\documentclass[aps,prd,10pt,twocolumn,superscriptaddress,flushbottom,nobibnotes,longbibliography]{revtex4-1}

\usepackage{hyperref}
\usepackage{graphicx}
\usepackage{amsfonts,amsmath,amssymb,bm,bbm}
\usepackage{color,soul}
\usepackage{slashed}
\usepackage[toc,page]{appendix}

\hypersetup{
    pdfnewwindow=true,      
    colorlinks=true,       
    linkcolor=blue,          
    citecolor=blue,        
    filecolor=blue,      
    urlcolor=blue        
}

\begin{document}

\title{New Experimental Constraints in a New Landscape for Composite Dark Matter}

\author{Christopher V. Cappiello}
\email{cappiello.7@osu.edu}
\thanks{\href{http://orcid.org/0000-0002-7466-9634}{orcid.org/0000-0002-7466-9634}}
\affiliation{Center for Cosmology and AstroParticle Physics (CCAPP), Ohio State University, Columbus, OH 43210}
\affiliation{Department of Physics, Ohio State University, Columbus, OH 43210}

\author{J.~I. Collar}
\email{collar@uchicago.edu}
\thanks{\href{http://orcid.org/0000-0002-0650-0626}{orcid.org/0000-0002-0650-0626}}
\affiliation{Enrico Fermi Institute, Kavli Institute for Cosmological Physics, and Department of Physics, University of Chicago, Chicago, Illinois 60637}

\author{John F. Beacom}
\email{beacom.7@osu.edu}
\thanks{\href{http://orcid.org/0000-0002-0005-2631}{orcid.org/0000-0002-0005-2631}}
\affiliation{Center for Cosmology and AstroParticle Physics (CCAPP), Ohio State University, Columbus, OH 43210}
\affiliation{Department of Physics, Ohio State University, Columbus, OH 43210}
\affiliation{Department of Astronomy, Ohio State University, Columbus, OH 43210} 

\date{\today}


\begin{abstract}
Certain strongly interacting dark matter candidates could have evaded detection, and much work has been done on constraining their parameter space. Recently, it was shown theoretically that the scattering cross section for $m_\chi \gtrsim 1$ GeV pointlike dark matter with a nucleus cannot be significantly larger than the geometric cross section of the nucleus.  This realization closes the parameter space for pointlike strongly interacting dark matter.  However, strongly interacting dark matter is still theoretically possible for composite particles, with much parameter space open.  We set new, wide-ranging limits based on data from a novel detector at the University of Chicago.  Backgrounds are greatly suppressed by requiring coincidence detection between two spatially separated liquid-scintillator modules.  For dark matter ($v \sim 10^{-3}$c), the time of flight would be $\sim 2~\mu{\rm s}$, whereas for cosmic rays, it would be $\sim 2~{\rm ns}$.  We outline ways to greatly increase sensitivity at modest costs.
\end{abstract}

\maketitle


\section{Introduction}
\label{sec:intro}

The identification of dark matter (DM) is a forefront problem in physics, as DM makes up most of the mass in the universe but its particle properties are unknown~\cite{Jun95, WMAP03, Ber04, Clo06, Fen10, Pet12}. As DM is only known to interact gravitationally, we have measured only its mass density averaged on astronomical distance scales.  While the GeV-scale weakly interacting massive particle (WIMP) is a popular DM candidate, still quite allowed~\cite{Lea18}, other candidates from axions to primordial black holes, with masses spanning many orders of magnitude, are also allowed~\cite{Pre82, Cov01, CAST08, ADMX09, Wan09, Bir16, Car16, Car17, ADMX18, Mon19}.

An important test of DM is its scattering cross section with nuclei. Direct-detection experiments are sensitive down to extremely small cross sections.  In this limit, the coherent spin-independent cross section of DM with a nucleus is related to that with a nucleon by the scaling relation $\sigma_{\chi A} = A^2\,(\mu_{\chi A}/\mu_{\chi N})^2\sigma_{\chi N}$, where $\mu_{\chi A}$ and $\mu_{\chi N}$ are the DM-nucleus and DM-nucleon reduced masses.  (For large $m_\chi$, this becomes $\sigma_{\chi A} \simeq A^4 \sigma_{\chi N}$.) But what are the {\it largest} cross sections that can be probed by direct-detection experiments?

\begin{figure}[t]
\centering
\includegraphics[width=\columnwidth]{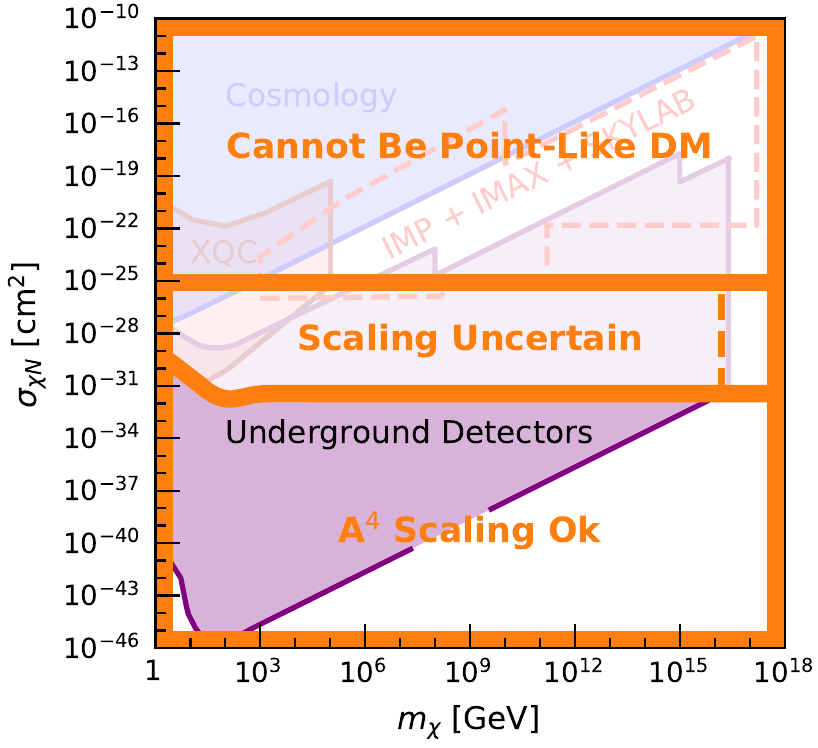}
\caption{Parameter space for spin-independent pointlike DM scattering, with previously claimed constraints overlaid by the new results of Ref.~\cite{Dig19} (we reproduce their Fig.~1).  In the middle region, pointlike DM is allowed but the constraints must be recalculated.  In the upper region, pointlike DM is not possible and the constraints are invalid.  Strong interactions remain possible for composite DM, our focus below.}
\label{fig:intro}
\vspace{-0.25cm}
\end{figure}

Until recently, the maximum cross section to which a direct-detection search is sensitive --- the so-called ceiling --- was computed based on DM attenuation in the atmosphere and Earth~\cite{Sta90, Col93, Alb03, Foo14, Kou14, Emk17b, Kav17, Mah17, Sid18}.  If the DM cross section with nuclei is too large, the DM will not reach the detector with enough energy, or at all.  This has been modeled either analytically or numerically, typically assuming the aforementioned scaling relation.  However, Ref.~\cite{Dig19} showed that this scaling fails for large cross sections, due to the breakdown of the first Born approximation, and even more significantly, that pointlike DM with a contact interaction cannot generally have a cross section much larger than the geometric size of the nucleus.  (For pointlike DM and the lightest nuclear targets, model-dependent $s$-wave resonances with cross sections above this limit may be broad enough to significantly affect the event rate \cite{Dig19}, but these targets are typically not relevant.)  This invalidates most prior calculations of the ceiling, and more generally of the excluded regions for high-cross section DM (e.g., Refs.~\cite{Alb03, Mac07, Khl07, Eri07, Khl08, Far17, Kav17, Mah17,  Bho18b, Bra18, Emk18, Far20}).  Figure~\ref{fig:intro} provides an overview.

However, large cross sections remain possible for composite DM, for which the open parameter space is large (see, e.g., Refs.~\cite{Sta90, Jac14, Gra18, Sid18}).  The properties of composite DM candidates are obviously more model-dependent than those of pointlike DM candidates.  One can make the simple but reasonable assumption that the DM particle is opaque to nuclei, with a scattering cross section equal to the DM particle's geometric size, regardless of the nuclear target~\cite{Sta90, Jac14, Gra18, Sid18}. As shown in Ref.~\cite{Dig19}, this is roughly accurate in the limit of strong coupling.  (Note that in this limit, the usual factors that relate the DM-nucleus and DM-nucleon cross sections --- the $A^2$ for coherence and the $\mu^2$ for the reduced mass squared --- are no longer present, so that the DM-nucleon cross section is equal to any DM-nucleus cross section.)  We view this as the most model-independent approach and conservative in a sense explained in Sec.~\ref{sec:results_results}; in the context of specific models, other possibilities could be explored. For elastic scattering of either pointlike or composite DM via a contact interaction, the total cross section is velocity-independent, although the kinematic range of the differential cross section does depend on velocity. Using the new framework discussed below, it would be interesting to renanalyze prior constraints on pointlike DM, but now for composite DM, but this is beyond our scope.

In this paper, we derive new limits on composite strongly interacting DM using a novel detector operated at shallow depth at the University of Chicago. This setup, previously used to set limits on sub-GeV moderately interacting DM~\cite{Col18, carlos}, consists of two liquid-scintillator modules separated by 50 cm, one directly above the other.  As we assume that the DM cross section is independent of the target nucleus, using a hydrogen-rich target material is ideal for a composite DM search because it maximizes the number of target nuclei per detector mass. Using hydrogen also allows us to probe DM-nucleon scattering directly, making it straightforward to translate our results to other models. For the large masses and cross sections we probe, DM passing through both modules would interact many times in each with minimal change in direction, leading to a coincidence signal with time separation $\sim 2~\mu{\rm s}$, due to the low RMS velocity, $\sim 10^{-3} ~c$, of DM in the standard halo model (SHM)~\cite{Lew96, Cho13, Gre17}.  Backgrounds due to cosmic rays and relativistic secondaries can also trigger both modules, but the time separation for these is prompt, $\sim 2$~ns. Fast neutrons can cause nuclear recoils in both modules with larger time separations, mimicking a DM signal.  As shown empirically below, the total background rate is very low.  The origins of the backgrounds are discussed further in Ref.~\cite{Col18}.

In Sec.~\ref{sec:setup}, we describe our experimental setup, data collection, and relevant backgrounds.  In Sec.~\ref{sec:results}, we analyze our data and report our constraints.  In Sec.~\ref{sec:future}, we detail ways to probe additional parameter space with new experiments. In Sec.~\ref{sec:conclusions}, we summarize our conclusions.


\section{Experimental Setup and Data}
\label{sec:setup}

The detector setup at the University of Chicago employs  two low-background EJ-301 modules~\cite{eljen}, each containing 1.5 liters of xylene-based, hydrogen-rich  liquid scintillator (with density 0.874 g/cm$^3$ and $4.82 \times 10^{22}$ H atoms/cm$^3$), to provide new experimental sensitivity to high-mass, strongly-interacting DM particles.  Using EJ-301 allows us to discriminate between events arising from interactions involving electron recoils (ER), like those produced by gammas and minimum-ionizing particles, and those involving nuclear recoils (NR), as expected  from the elastic scattering of neutral particles, be those fast neutrons, or the sought-after DM candidates. This discrimination ability arises from the dissimilar scintillation decay constants for ERs and NRs in EJ-301, the second favoring delayed emission~\cite{decays}. 

For particles producing signals in both modules, we exploit the time-of-flight (TOF, $\Delta$t) between the modules.   Slow-moving DM candidates obeying the kinematics expected from the SHM would generate highly characteristic signatures in $\Delta$t~\cite{Col18}. The methodology, experimental arrangement, detector calibrations, environmental backgrounds, as well as the active and passive shielding surrounding the modules in the shallow underground laboratory (6.25 m.w.e.), are described in full detail in Ref.~\cite{Col18}. Dedicated runs using this setup were performed for the present search.

\begin{figure}[t]
\centering
\includegraphics[width=\columnwidth]{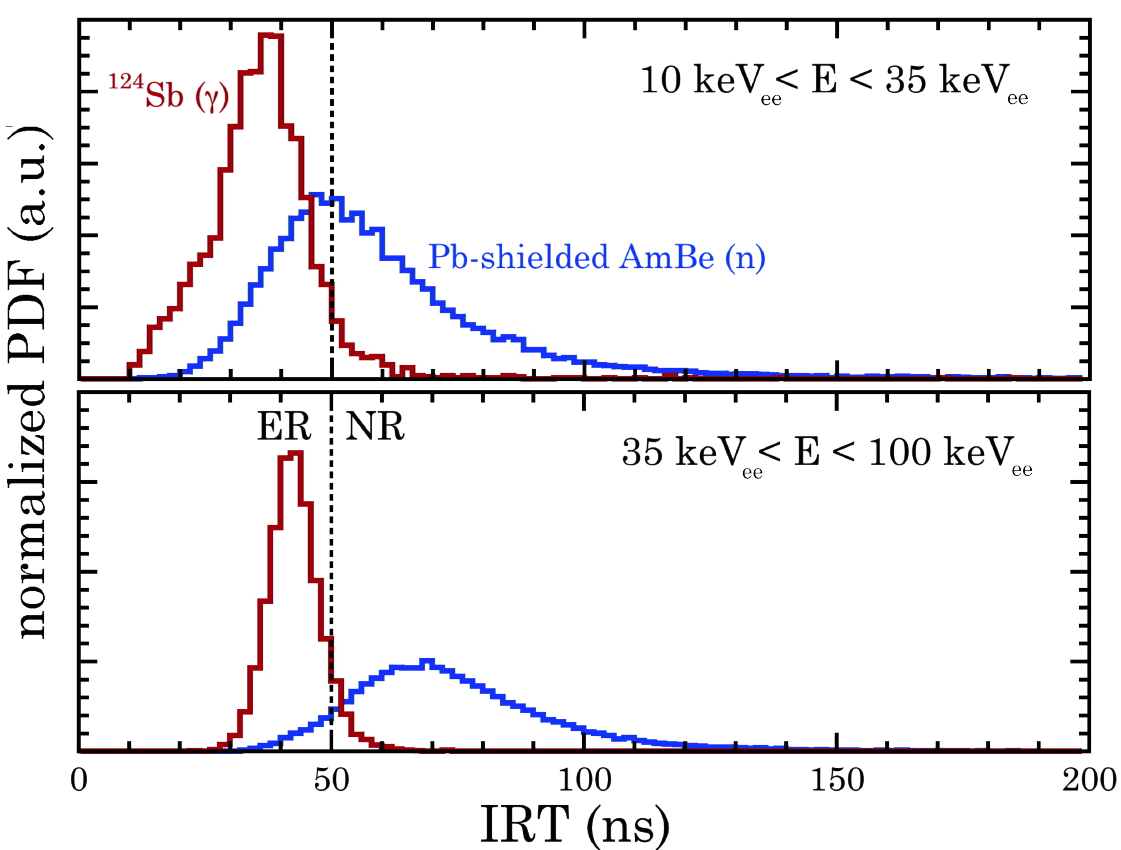}
\caption{Separation between ER-like (red) and NR-like (blue) depositions in individual EJ-301 modules, measured using gamma and neutron sources via the IRT method. See Ref.~\cite{Col18} for further details. This separation improves with increasing energy; above 100 keV$_{ee}$, it remains nearly identical to that shown in the bottom panel.} 
\label{fig:irt}
\end{figure}

Figure~\ref{fig:irt} shows the results of calibrations to measure the separation between ER-like and NR-like events.  The energy calibration of the modules was obtained using full energy depositions from $^{241}$Am 59.5 keV gammas,  the Compton edge from 2,091 keV $^{124}$Sb gammas, and distinct muon traversals along the vertical axis of the detectors, which deposit approximately 22 MeV in organic liquid scintillator modules of this size. NR calibrations are obtained using neutrons from a Pb-shielded AmBe source. Photomultiplier gain was reduced with respect to that employed in Ref.~\cite{Col18}, allowing for the detection of electron-equivalent (ee) energies ranging from 10 keV$_{ee}$ up to approximately 25 MeV$_{ee}$.  For events depositing energies above 100 keV$_{ee}$, the promptest fraction of the scintillation pulse saturates the range of the 8-bit digitizer employed. This precluded a search for the characteristic pulse-shape distortion expected from a slow-moving particle losing energy continuously within a scintillating medium. However, this did not affect our ability to efficiently discriminate between particles losing energy via ERs or NRs, achieved over the full energy range via a modified integrated rise-time (IRT) method~\cite{luo, ronchi}. (For further details, see Ref.~\cite{Col18}.) Similarly, an ad-hoc modification of the energy scale above 100 keV$_{ee}$ was used to determine the magnitude of energy depositions partially saturating the digitizer.  For purposes of extracting limits, the calculated nuclear recoil energies imparted by heavy DM were converted to electron-equivalent energies using a model of the quenching factor for proton recoils in EJ-301, described and validated in Ref.~\cite{awe}.

The measured compound efficiency for identifying NR-like depositions in both modules (IRT $>$ 50 ns in each) is nominally 36.9\% for events with 10--35 keV$_{ee}$ in each module, and 85.9\% for higher energies (Fig.~\ref{fig:irt}). However, these efficiencies are conservative lower limits, as DM particles continuously losing energy via NRs would produce larger IRT values than the neutron scatters used in these calibrations, a result of their $\sim 300$-ns TOF through each module. In our analysis, we conservatively adopt these lower limits, with the remaining fractions of events being misidentified as electron-recoil events and rejected.

\begin{figure}[t]
\centering
\includegraphics[width=\columnwidth]{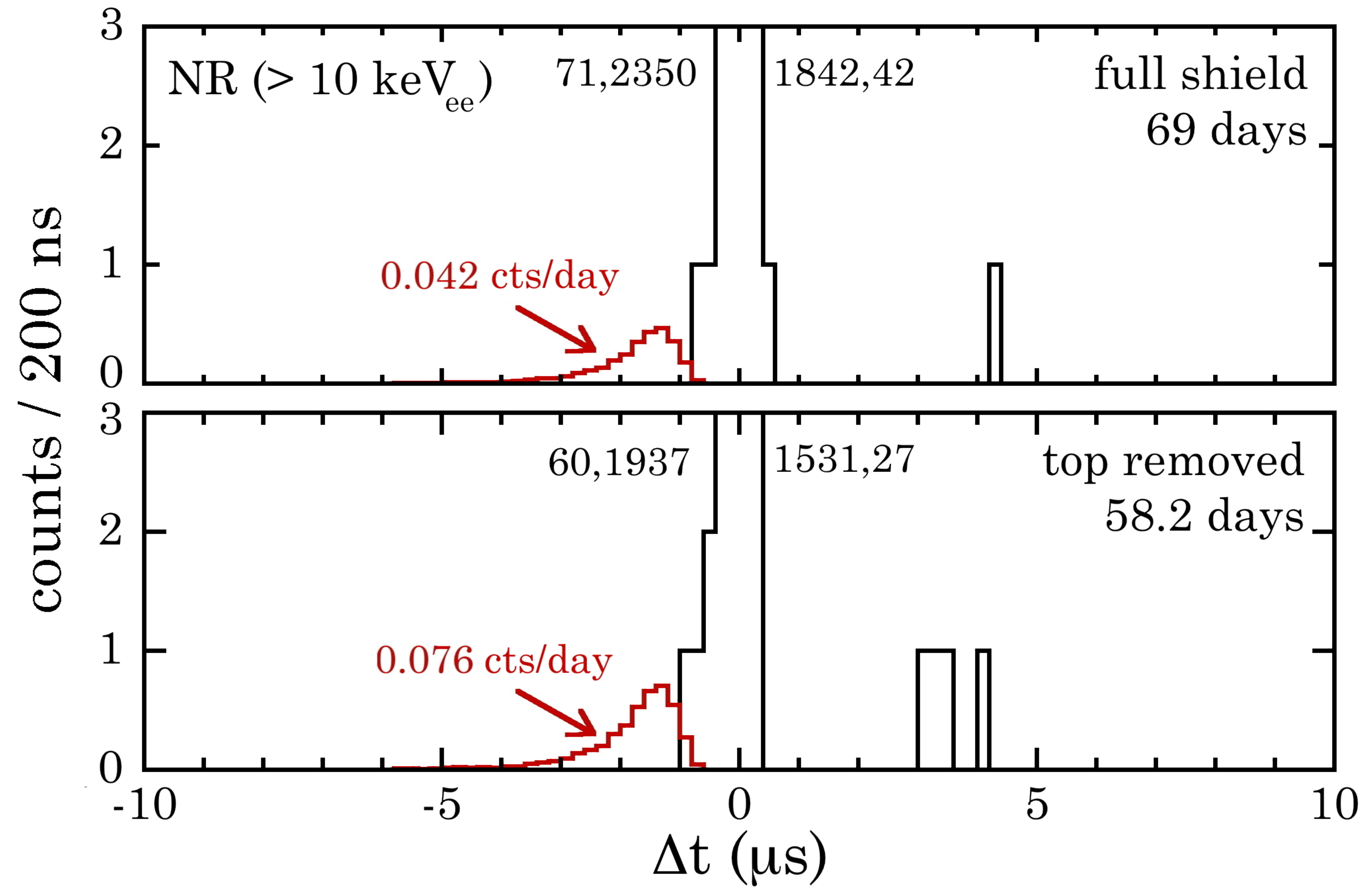}
\caption{TOF distributions for events above 10 keV$_{ee}$ collected during the two runs. As the heights of the four central bins (corresponding to prompt coincidences between modules) are outside the vertical range of plot, we note their values.  Red histograms show the maximum contributions from DM candidates (which are slow-moving) allowed by the data. Upward-moving DM particles are assumed to be stopped by the Earth, limiting our search to negative TOF values~\cite{Col18}.} 
\label{fig:expresults}
\end{figure}

Figure~\ref{fig:expresults} displays the distribution in $\Delta$t of events with NR-like scintillation deposits of greater than 10 keV$_{ee}$ in each module.  In the first run, 69 days long (top panel), the lead shield entirely surrounding the modules was as described in Ref.~\cite{Col18}. In the second run, 58.2 days long (bottom panel), six lead bricks were removed from the top of the shield. In this case, all but a negligible fraction of straight trajectories traversing both modules go through the hexagonal opening created by this action, removing the attenuation caused by 15 cm of lead.  The spike of events with $\lvert  \Delta t \rvert < 1~\mu{\rm s}$ are prompt coincidences induced by relativistic particles; we exclude this background via a time cut.  

We searched for DM particles with vertically downward trajectories through both detectors (as in Ref.~\cite{Col18}).  The region of interest (ROI) is $-3.5~\mu{\rm s} < \Delta t < -0.8~\mu{\rm s}$, corresponding to the velocity distribution of dark matter particles in an Earth-bound laboratory \cite{Col18}.  In the first and second runs, the numbers of viable candidate events in the ROI were zero and one, respectively (Fig.~\ref{fig:expresults}). This ROI corresponds to DM velocities between approximately 140 and 625 km/s, which includes over 90\% of the DM velocity distribution at Earth (that is, when the Sun's velocity through the DM halo is included). The $\lesssim$10\% inefficiency has a negligible effect on our results.

The mean expected background counts in this $\sim\!2.7~ \mu{\rm s}$-wide ROI can be calculated based on the counts with $\lvert \Delta t \rvert > 1~\mu{\rm s}$, keeping in mind that random coincidences between uncorrelated energy depositions in each module give rise to a flat distribution of events~\cite{Col18}.  (This wide range includes the narrow range of the ROI, but the Feldman-Cousins approach allows us to separate signal and background by comparison.)  In the first and second runs, the numbers of such events were one and four, respectively.  The background counts expected in the ROI are thus  1 event/18 $\mu{\rm s} \times 2.7 \mu{\rm s} = 0.15$ events for the first run, and 4 events/18 $\mu{\rm s} \times 2.7~ \mu{\rm s} = 0.6$ events for the second (the 18-$\mu$s factor is the 20-$\mu$s width of the TOF search window minus the range $\lvert \Delta t \rvert < 1~\mu{\rm s}$ that is dominated by relativistic-particle backgrounds).

Using a standard Feldman-Cousins approach~\cite{feldman}, these background expectations and the observed numbers of viable candidates can be translated into 90$\%$ confidence level (C.L.) intervals for the maximum number of DM candidates allowed by the datasets, yielding 2.9 and 4.4 events, respectively.  Their expected characteristic TOF distributions~\cite{Col18} and the maximum daily rates of DM interactions allowed by the data are shown in Fig.~\ref{fig:expresults}. 


\section{Analysis and Results}
\label{sec:results}

Following the assumption above that DM is opaque to nuclei, we compute the event rate in the detector as a function of mass and cross section.  To trigger the detector, the DM must scatter many times in each module.


\subsection{Incident Dark Matter Rate}

We assume that DM has the bulk properties predicted by the SHM, with a Maxwellian velocity distribution, a velocity dispersion of 270 km/s, and a local density of 0.3 GeV/cm$^3$~\cite{Lew96, Cho13, Gre17}. The impact of altering these assumptions is minimal, as discussed below.

We consider only DM particles arriving from above and passing through both detector modules (Earth is opaque to strongly interacting DM). Specifically, we require that a DM particle reaching the center bottom of the lower module pass through the top of the upper module. The cylindrical modules are 10 cm in height, 10 cm in diameter, and have 50 cm of space between them, so this requirement means the experimental setup is sensitive to a fraction $1.3 \times 10^{-3}$ of $4\pi$, and we thus accept this fraction of the total incoming DM flux. While this solid angle is small, for large enough cross sections, every DM particle passing through both modules would interact, so the event rate can still be high. If there were no attenuation above the detector and  downgoing DM particles that passed through both modules triggered the detector, the rate would be $10^{11}\,({\rm GeV}/m_{\chi})$/day.  This is lowered by more realistic assumptions, that we calculate below.


\subsection{Attenuation}

DM reaching the detector must pass through about 10 meters of water equivalent (m.w.e\.) of atmosphere, as well as 6 feet ($\sim$6.25 m.w.e.) of concrete shielding above the laboratory. We model attenuation by assuming that DM particles travel along straight-line trajectories, with each particle suffering the average energy loss in each collision ($\cos\theta = 0$ in the CM frame), taking into account the loss of energy for the DM particle as it propagates.  This formalism is widely used for computing DM attenuation, e.g. in Refs.~\cite{Sta90, Alb03, Kou14, Emk17b, Kav17, Sid18}, and is an excellent approximation for heavy DM.  The energy loss rate is then
\begin{equation}
\frac{dE}{dx} = - \sum_j n_j \sigma_{\chi} \langle \Delta E_j \rangle\,,
\end{equation}
where the sum is over different nuclei, $n_j$ is the number density of the $j$th nuclear species, and $\langle \Delta E_j \rangle$ is the average energy loss in a collision with a nucleus of species $j$. For $m_{\chi} \gg m_j$, and $E$ being the initial DM kinetic energy, the average energy change per scatter is
\begin{equation}\label{DE}
\langle \Delta E_j \rangle \simeq 2 E \, \frac{m_j}{m_{\chi}}\,.
\end{equation}
From the above equations, and because the cross section does not depend on the nuclear species due to being set by the geometric size of the DM, the energy loss depends only on the total mass column density of nuclei encountered by the DM, and not the elemental composition. Heavier nuclei have lower number density per unit mass density (by $1/A$), but also have more energy loss per collision (by $A$), such that the total stopping power depends just on the mass column density. Thus, different from the usual case, a given mass column density of lead does not have significantly higher stopping power than the same mass column density of concrete.

Approximating the energy loss as equal for all DM particles is well motivated for the high cross sections we consider. The DM particles reaching the detector would scatter hundreds or thousands of times while traversing the atmosphere on near-vertical trajectories, meaning the fractional 1$\sigma$ Poisson fluctuation in the number of collisions is at most a few percent. Similarly, for the high masses of concern, it is an excellent approximation to treat particle trajectories as straight lines. For $m_{\chi} \gg m_j$, the DM lab-frame trajectory is deflected by an angle $\theta \sim m_j/m_{\chi}$ in one scattering~\cite{Kav16}. As we show below, our analysis is sensitive to $m_{\chi} \gtrsim 10$ TeV, so $\theta \lesssim 10^{-3}$ for typical nuclei in the overburden. The cumulative RMS deflection angle is then $\theta/\sqrt{N_{\rm scatt}} \lesssim 10^{-4}$, where $N_{\rm scatt} \sim 100$ is appropriate for the minimum cross sections of our allowed region.

The amount of kinetic energy a DM particle retains after propagation through the atmosphere and detector shielding is a decaying exponential in $\sigma_{\chi}$, so a small change in $\sigma_{\chi}$ leads to a large fractional change in the amount of energy loss. For this reason, although we take into account the energy dependence of attenuation, simpler calculations would give similar results. That is, defining the ceiling to be the cross section where DM loses $10\%$, $50\%$ or $90\%$ would all give similar results to our procedure, where we compute the energy a DM particle deposits in the detector even if it has lost nearly all its energy before reaching the detector. This is different from calculations for lower masses, where large scattering angles and fluctuations in the number of collisions make detailed propagation codes necessary \cite{Emk17c, DMATIS, Cap19b}.


\subsection{Signals in the Detector}

For interactions in the detector modules, we conservatively consider only DM scattering with protons.  With an RMS DM velocity of $\sim 10^{-3}c$, and DM much heavier than a proton, the typical individual proton recoil energy is $\sim 1$ keV (as can be seen from Eq.~\ref{DE} and derived elsewhere,  e.g., Ref.~\cite{Mac07}). The equivalent light yield is reduced from this by a multiplicative quenching factor, which we take to be $20\%$ based on the EJ-301 response model from Ref.~\cite{awe}. In fact, the quenching factor depends on the recoil energy, but the fit in Ref.~\cite{awe} shows that it is actually above $20\%$ for the vast majority of collisions we consider (recoil energies from about 0.15--4 keV), so our assumption is conservative. Including carbon recoils would have only a small effect on our results:  although carbon recoils have about 12 times more kinetic energy, their scintillation emission is much more quenched, with a quenching factor of $\sim$1\%~\cite{awe}.  Including carbon recoils would increase the total light yield in the detector by less than a factor of 2, which would hardly change the excluded region relative to the huge scale of the plot.

For a DM particle passing through the detector, we sum the calculated electron-equivalent energy of all individual proton recoils to compute the total scintillation signal. We require a minimum total energy deposition of 10 keV$_{ee}$ in each module, which corresponds to $\sim 50$ collisions.  It would take DM about 300 ns to pass through each module, which is comparable to the long component of the scintillation decay constant ($\sim 270$ ns) for EJ-301~\cite{eljen}, so we integrate the total deposited energy. As described in the previous section, we then use the NR-like particle identification and the time-of-flight criteria to discriminate between DM events and background events, ruling out DM parameter space that corresponds to too large of an event rate.  At the highest cross sections we consider, the distance between collisions becomes small enough that it is no longer a good approximation to sum the quenched energies of individual collisions, as the energy-deposition regions would overlap; however, in this limit, the energy deposited is vastly above threshold.


\subsection{Results}
\label{sec:results_results}

\begin{figure}[t]
\centering
\includegraphics[width=\columnwidth]{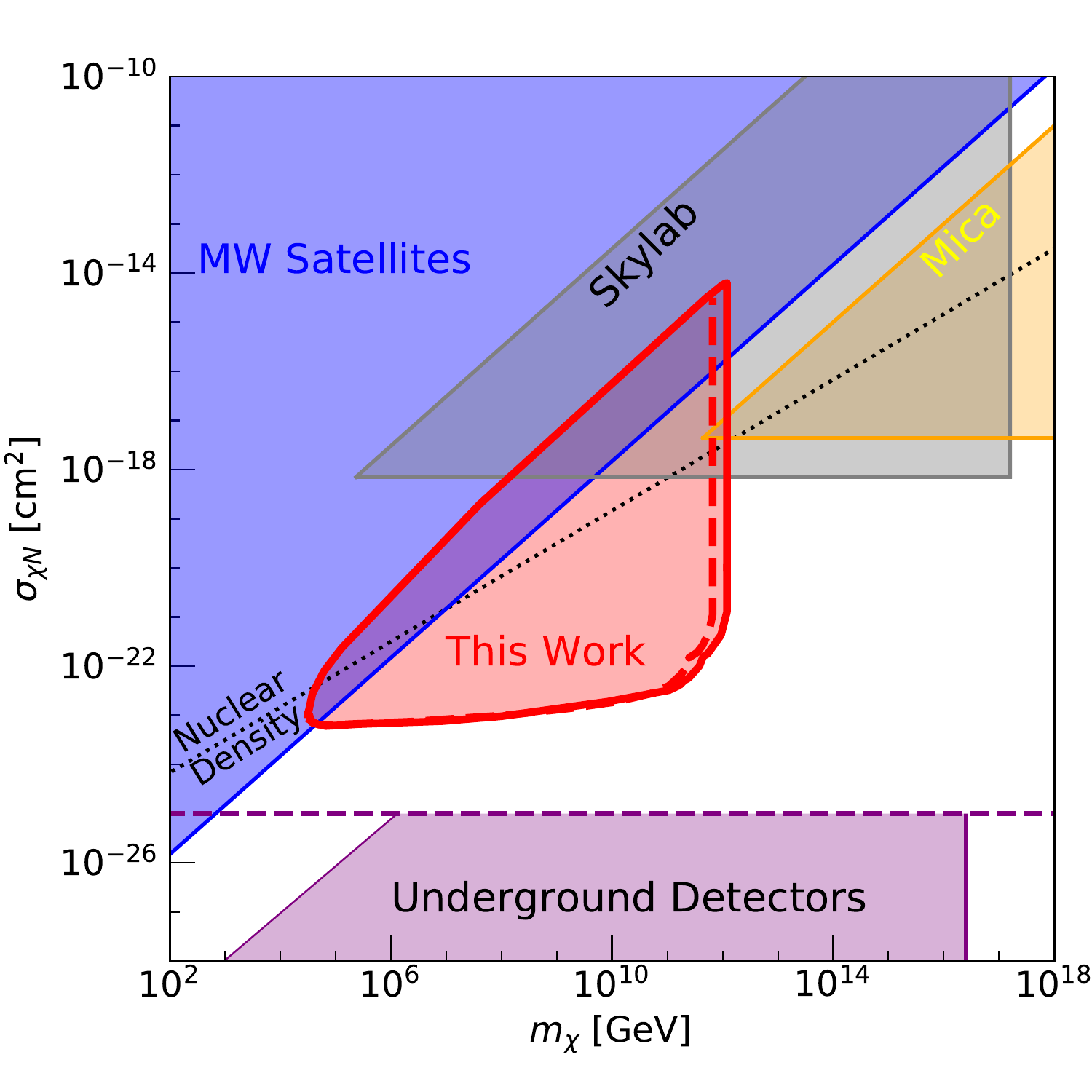}
\caption{Constraints (90\% C.L.) on composite DM.  The red solid boundary is the exclusion region from our search (for the dashed red line, see the text).  Also shown are other limits based on composite DM scattering with nuclei, in the Skylab satellite~\cite{Sta90} and in ancient underground mica~\cite{Pri86, Jac14}, plus one cosmological constraint based on how DM-proton scattering would affect the structure formation of Milky Way satellites~\cite{Nad20}.  The usual limits on pointline DM from underground detectors are shown, but we cut them off by the maximum allowed DM-nucleon cross section~\cite{Dig19}.  The dotted line shows, just for comparison, where the internal density of the DM would be comparable to that of typical nuclei.}
\label{fig:results}
\end{figure}

Figure~\ref{fig:results} shows our constraints on composite DM. The red triangular region (``This Work") is ruled out by the null results of our search. Parameter space to the left of the dashed line is excluded by both the run with lead shielding and the run without it. Parameter space to the right of the dashed line is excluded only by the run with full shielding. Because the run with shielding had fewer observed events, we can exclude slightly more massive (and thus lower flux) DM. The slight indentation in the bottom-right corner is due to the aforementioned difference in discrimination efficiency for events above and below 35 keV$_{ee}$: DM particles depositing more than 35 keV$_{ee}$ are more efficiently distinguished from electron-recoil events than lower-energy DM particles, so at higher cross section our analysis is sensitive to slightly lower flux (and thus higher mass). 

The constraints we show are on the geometric size of an opaque, composite DM particle.  They should not necessarily be thought of in the usual framework of spin-dependent and spin-independent scattering.  (In any case, we directly probe the DM-proton cross section.)  Those terms typically refer to specific operators used in nonrelativistic effective field theory, where many other operators are possible~\cite{Fit12}. We consider a DM particle with physical size much larger than the wavelength associated with the momentum transfer of the scattering, so the nucleus does not ``see" the entire DM state.  However, we assume that the cross section saturates to the geometric size of the DM particle due to strong internal couplings of the DM composite state.  For increasing DM size, there may be a model-dependent (e.g., see Refs.~\cite{App13, Har15, Chu18}) form factor that prevents this saturation.  However, the moderate cross sections at the bottom edge of our exclusion region are already sufficient to cause an excess rate, and the rate nominally increases linearly with increasing cross section, so the rate should be quite high even with a form factor.  In addition, with a form factor the diagonal edge could become much higher, as it has a strong exponential dependence on the cross section.  In light of this, we view our assumptions as conservative.

The underground-detector limits, derived assuming that $\sigma_{\chi A} \simeq A^4 \sigma_{\chi N}$, are shown cut off at the maximum allowed DM-nucleon cross section for typical nuclear targets~\cite{Dig19}, though even below this line their limits may be improperly derived. For pointlike DM, cross sections above this limit can be obtained for $s$-wave resonances, but these model-dependent resonances are quite narrow in energy range for all but the lightest nuclei, and are not usually considered for direct detection experiments. 

Our limits on composite DM should not be directly compared to limits on pointlike DM, but we show these pointlike limits for orientation. Really, the results from underground direct-detection experiments should be reanalyzed for composite DM both below this line (where the scaling is uncertain) and above it (where there may still be some sensitivity).  The underlying issue is that those analyses used a DM-nucleus cross section that was far too large due to using the coherent scattering scaling relation between DM-nucleus and DM-nucleon cross sections.  However, instead using just the geometric cross section of the DM may still allow signal rates that are detectable above backgrounds.  New work is needed to determine the actual exclusion regions for composite DM.  The Skylab and mica results were explicitly reported as limits on composite DM, also assuming that the DM is opaque to nuclei~\cite{Sta90, Pri86, Jac14}. They are more directly constraints on $\sigma_{\chi A}$, but were recast as limits on $\sigma_{\chi N}$ under the assumption that $\sigma_{\chi A}$ = $\sigma_{\chi N}$ = $\sigma_{\chi}$, the geometric size of the DM particle. The cosmological constraints are derived by considering only DM-proton scattering~\cite{Nad20}.

The right edge of our region is set by the DM density; beyond the edge, the number density of DM ($\rho_\chi/m_\chi$) becomes so low that there are simply no signal events within this exposure. The maximum mass we are sensitive to is roughly $m_\chi \simeq \rho v (A \Omega t) \simeq 4 \times 10^{12}$ GeV, where the latter factor is the total exposure, with $A$ being the cross sectional area of the top of the detector and $\Omega$ being the solid angle of acceptance. The diagonal edge is set by attenuation above the detector; beyond the edge, the flux of DM with any appreciable energy (or at all) vanishes exponentially. This is roughly determined by the cross section for which the total mass of scattered nuclei is comparable to the DM mass: $\sigma_{\chi} n_A m_A d\simeq m_{\chi}$, where $n_A$ is the average number density of nuclei of species $A$ and $d$ is the distance traveled through the atmosphere or shielding. With the atmosphere being about 10 m.w.e., we get $\sigma_{\chi}/m_{\chi} \simeq 10^{-27}$ cm$^2$/GeV. The bottom edge is set by the detector threshold; beyond the edge, the DM collisions are too few to deposit enough energy to trigger the detector at that threshold (which is set by the background rates). This is determined by the requirement that the DM scatter at least 50 times with protons in each module, so $\sigma_{\chi} \simeq 50/(n_H L) \simeq 10^{-22}$ cm$^2$, with $L$ being the length of a detector cell and $n_H$ being the proton density in the cell. The bottom edge is not flat, as explained in the next paragraph. Within the excluded region, the calculated DM event rates are generally enormous.  For example, for a point near the center (mass $10^9$ GeV and cross section $10^{-20}$ cm$^2$), every DM particle would deposit about 8 MeV in each module and the rate of such depositions would be about 10$^4$/day.  

The bottom edge is not flat due to Poisson fluctuations.  To trigger the detector, DM must deposit 50 keV in each module to produce 10 keV$_{ee}$ of scintillation light, taking into account the 20\% quenching factor.  With an average proton recoil energy of 1 keV, this means a DM particle must scatter $\sim$50 times in each module.  At the right end of the bottom edge, where the flux of DM is smallest, the minimum excluded cross section is determined by the cross section for which the Poisson expectation is indeed about 50 collisions.  The exact value is calculated by requiring a certain number of signal events in the exposure, following the details given above.  At lower mass, however, the flux increases as $1/m_\chi$, giving vastly more trials, making it probable that a smaller cross section and Poisson expectation can still give a Poisson yield of 50 or more collisions.  At the left end of the bottom edge, roughly $10^8$ DM particles pass through both modules in $\sim$60 days, allowing a factor of a few improvement in the cross section (i.e., a Poisson expectation of $\sim$20 collisions is enough).

Our results are stable to deviations within uncertainties of the assumed bulk properties of DM.  A change in the DM density would affect only the right edge, and would not be visible on the scale of the figure.  A change in the DM RMS velocity would primarily affect the bottom edge, but only slightly.  A change in the shape of the DM velocity distribution for the same RMS velocity would affect primarily the diagonal edge, but again only slightly.  Thus changes in the DM bulk properties like those discussed in Refs.~\cite{Lis11, Mao14, Boz16, Slo16, Nec18, Nec18b, Eva18, OHa19, Bes19}, the subject of much recent work, can be neglected.  We assumed that DM arrives isotropically, which is not strictly true, as it arrives preferentially from the direction in which the Sun is moving around the galaxy (the so-called WIMP wind, e.g., Refs.~\cite{Sci09, Kav17}). However, the overhead direction in Chicago actually points into the WIMP wind for part of the day, making this assumption conservative \cite{Sci09}.  Because changing the incoming flux by even a factor of a few would not visibly affect Fig.~\ref{fig:results}, we neglect this effect.

For future work, we note some considerations that may become important eventually. Near the top right of our exclusion region, the diameter of the DM particle becomes comparable to interatomic spacing, meaning the DM may scatter with multiple nuclei at the same time.  While this would not change the physical deposited energy, it could change the quenched energy.  However, in this region of parameter space, the energy depositions are vastly above threshold.  Near the bottom right of our exclusion region, below the line $\sigma_{\chi}/m_{\chi} \simeq 10^{-33}$ cm$^2$/GeV, DM particles may pass through Earth, allowing upgoing events, which would only increase the signal rate.


\section{Extensions for Future Work}
\label{sec:future}

What would be needed to significantly expand coverage of the composite DM parameter space?  We first focus on how a detector like ours could be improved, and then consider other options.

The diagonal edge of the sensitivity region cannot be substantially improved for detectors near Earth's surface, as the DM flux is exponentially depleted by attenuation.  (Some improvement should be possible by using a better digitizer for larger pulses, which would enable new capabilities to recognize slow-moving tracklike events.)  The right edge, in contrast, could be greatly improved.  In the setup above, the top area of each detector was $\sim 80$ cm$^2$, the runtime was $\sim 0.2$ yr, and the solid-angle penalty was $\sim 10^{-3}$.  It is easy to imagine, as discussed below, how the exposure could be improved by several orders of magnitude.  If the backgrounds can be kept near-negligible, this would improve the sensitivity of the right edge by the same factor.  For the bottom edge, there is room to improve by lowering the threshold energy, which would require reducing backgrounds.  This would be especially important for covering the gap between our sensitivity region and that of conventional underground detectors.  Reanalyses of the latter taking into account the breakdown of the usual scaling relation would also help.

A potential setup could consist of a horizontal checkerboard array of modules, with this array replicated at multiple heights.  This would increase the area and solid angle exposure, and of course one could choose to run for a longer time.  With modules at more than two heights, the required multiple coincidences would reduce uncorrelated backgrounds, which could allow a lower threshold per module. Greatly improved sensitivity could be achieved for moderate costs.  Similar detectors probing lower cross sections may be sensitive not just to downgoing DM, but also to upgoing DM, in which case the angle-dependent attenuation in Earth must be treated properly and the sign of the time delay accounted for, though this would increase the signal rate by at most a factor of two.

Other setups could help cover more of the parameter space.  Significantly moving the diagonal edge would require a satellite or high-altitude balloon experiment.  The column density of the overburden could realistically be reduced by about $\sim 10^3$, which would increase the sensitivity by the same factor.  (For example, note that the Skylab diagonal edge is about this much higher than ours.)  Our exclusion region is already partially covered by the cosmological constraints~\cite{Nad20}, though direct experimental sensitivity is always preferred.  Another low-threshold surface detector, a special run of CRESST~\cite{Ang17}, should be reanalyzed for composite DM, as should DAMA's search for strongly interacting DM~\cite{Ber99}, and constraints from planetary heat flow~\cite{Mac07, Bra19} or heating of celestial bodies~\cite{Kou10, Bar17, Bel19, Das19, Bel20, Das20}.

It may be fruitful to consider large underground detectors, as done in Ref.~\cite{Bra18b}.  While their diagonal edges could be as much as $\sim 10^2$ times lower than ours (for detectors at thousands of m.w.e.\ depth), those could still nearly meet the cosmological constraints~\cite{Nad20}.  Such detectors would have the advantage of a huge exposure.  In the ideal situation, a highly segmented large detector would be an expanded version of the stacked arrays of modules discussed above. Data from other detectors with tracking capability, such as the proposed MATHUSLA~\cite{Lub19}, could be analyzed for such a search. It may be possible to effectively achieve imaging of tracklike events in homogeneous detectors with excellent position resolution, such as the time-projection chambers used in many large DM experiments~\cite{Akerib_2017, Aprile_2017, Calvo_2017, Agnes_2018, Akerib_2020}, bubble chambers like PICO-60~\cite{Amo17} or perhaps even DUNE~\cite{abi2020deep}. Also potentially sensitive to this parameter space are certain sub-detectors of the MACRO experiment~\cite{Amb02}. In some other detectors such as Borexino~\cite{Alimonti_2009} or SNO+ \cite{And15}, the isotropized scintillation light usually makes directional reconstruction difficult, but that changes for DM particles that leave long, distinctive tracks that develop $\sim 10^3$ times more slowly than muon tracks~\cite{bramante_2019}.


\section{Conclusions}
\label{sec:conclusions}

To discover DM, we must search broadly.  In an interesting class of models, DM has evaded detection by interacting too strongly with ordinary matter, instead of too weakly, as commonly assumed.  The parameter space of pointlike strongly interacting DM has largely been eliminated by the theoretical considerations of Ref.~\cite{Dig19}, superseding much prior work~\cite{Alb03, Mac07, Khl07, Eri07, Khl08, Far17, Kav17, Mah17,  Bho18b, Bra18, Emk18, Far20}.  However, it also highlights the open parameter space for composite strongly interacting DM~\cite{Sta90, Jac14, Gra18, Sid18}, for which prior results on pointlike DM could be reanalyzed.

In this paper, we present experimental tests of  composite DM based on a dedicated search using a novel detector at the University of Chicago.  This detector, while small and near the surface, has powerful sensitivity, covering large regions of the parameter space that had not yet been probed.  We did not find DM candidates, and accordingly set limits.  Our analysis shows that for composite DM, terrestrial detectors can probe cross sections above cosmological constraints~\cite{Nad20}, reducing the need for rocket- or space-based experiments. At moderate costs, an improved version of our detector could have greatly improved sensitivity, potentially extending significantly lower in cross section and several orders of magnitude higher in DM mass.

We have also detailed how progress could be made more generally, so that strongly interacting composite DM could be probed over a much wider parameter space.  It is important to do this systematically.  While a strongly interacting DM particle might seem unlikely relative to typical theoretical prejudices, those prejudices have also not led us to any discovery of DM.  A strongly interacting DM particle could also enable a host of new laboratory studies.  New analyses of existing data will be an important part of this, and we encourage them.


\section*{Acknowledgments}

We are grateful to Carlos Blanco, Joseph Bramante, Matthew Digman, Timon Emken, Brian Fields, Chris Hirata, Bradley Kavanagh, Jason Kumar, Ben Lillard, Annika Peter, Nirmal Raj, Anupam Ray, Juri Smirnov, and Xingchen Xu for helpful comments and discussions. We also thank the anonymous referee for helpful comments that improved the paper. CVC and JFB were supported by NSF grant Nos.\ PHY-1714479
and PHY-2012955.  JIC was supported by NSF grant No.\ PHY-1506357.


\bibliography{CompositeDM}

\end{document}